\documentclass[preprint,noshowkeys,amsmath,amssymb,superscriptaddress,nofootinbib]{revtex4-1}

\usepackage{bm}
\usepackage{slashed}
\usepackage[version=3]{mhchem}
\usepackage{ulem}
 
\usepackage[pdftex]{graphicx,color}
\usepackage{epstopdf}
\usepackage[bookmarks=true,bookmarksnumbered=true,bookmarkstype=toc, colorlinks=true, 
citecolor=blue, linkcolor=blue ,urlcolor=blue]{hyperref} 
\usepackage{units}

\newcommand{\lsim}{\raise0.3ex\hbox{$\;<$\kern-0.75em\raise-1.1ex\hbox{$\sim\;$}}}
\newcommand{\gsim}{\raise0.3ex\hbox{$\;>$\kern-0.75em\raise-1.1ex\hbox{$\sim\;$}}}
\usepackage{mathtools}

\renewcommand{\eqref}[1]{Eq.~(\ref{#1})}

\usepackage{color}

\definecolor{green}{cmyk}{1,0,1,0}

\definecolor{pink}{cmyk}{0,0.5,0,0}

\definecolor{pastelpink}{cmyk}{0,0.25,0,0}

\definecolor{softpink}{cmyk}{0,0.125,0,0}

\definecolor{purple}{cmyk}{0.5,1.0,0.1,0}

\definecolor{violet}{cmyk}{0.75,1,0.25,0}

\usepackage{ulem}  

\newcommand{\lmlt}{L_\mu - L_\tau}

\usepackage[utf8]{inputenc}

\begin{document}

\preprint{UME-PP-019}
\preprint{STUPP-21-248}
\preprint{KYUSHU-HET-227}

\title{Electron Beam Dump Constraints on Light Bosons \\
with Lepton Flavor Violating Couplings}

\author{Takeshi Araki}
\email{t-araki@den.ohu-u.ac.jp}
\affiliation{Faculty of Dentistry, Ohu University, 31-1 Sankakudo, Tomita-machi, Koriyama, Fukushima 963-8611,  Japan}

\author{Kento Asai}
\email{asai@krishna.th.phy.saitama-u.ac.jp}
\affiliation{Department of Physics, Faculty of Science, Saitama University, Sakura-ku, Saitama 338--8570, Japan}

\author{Takashi Shimomura}
\email{shimomura@cc.miyazaki-u.ac.jp}
\affiliation{Faculty of Education, University of Miyazaki, 1-1 Gakuen-Kibanadai-Nishi, Miyazaki 889-2192, Japan}
\affiliation{Department of Physics, Kyushu University, 744 Motooka, Nishi-ku, Fukuoka, 819-0395, Japan}

\begin{abstract}
We study constraints on light and feebly interacting bosons with lepton flavor violation 
from electron beam dump experiments. 
Scalar, vector, and dipole interactions of the bosons are analyzed, respectively, and 
excluded regions from the searches for decays into electron-positron pairs are derived.
It is found that parameter regions unconstrained by flavor violating decays of muon can be excluded using 
the results of the E137 experiment.
We also discuss the impact of the search for flavor violating decays of the light bosons in electron beam dump experiments.
\end{abstract}

\date{\today}

\maketitle

\section{Introduction}
\label{sec:introduction}

For the last decade, the searches for new physics beyond the Standard Model (SM) have been primarily led by the LHC with particular attention to the regions of TeV-scale mass with $\cal{O}$(1) coupling.
Despite a great deal of experimental and theoretical efforts has been made in that direction, no signals have been found so far.
Given the fact, recently, there is a growing interest in a neutral boson having sub-GeV mass and feeble interactions with the SM particles.
In fact, there are several theoretical motivations to consider such a feebly-interacting light boson; For instance, it plays a crucial role in 
the dark matter problem \cite{TuckerSmith:2001hy,Boehm:2003hm,Pospelov:2007mp,Izaguirre:2015yja,Knapen:2017xzo,Fitzpatrick:2020vba,Duerr:2020muu}, 
the muon $g-2$ anomaly \cite{Baek:2001kca,Pospelov:2008zw,Altmannshofer:2014pba}, 
the Hubble tension \cite{Escudero:2019gzq,Escudero:2019gvw,Escudero:2020dfa,Araki:2021xdk}, 
and it is also claimed that the existence of the light boson helps us understand the observed energy spectrum of high energy cosmic neutrino \cite{Ioka:2014kca,Ng:2014pca,Ibe:2014pja,Araki:2014ona,Kamada:2015era,DiFranzo:2015qea,Araki:2015mya,Mohanty:2018cmq,Carpio:2021jhu}.

Feebly-interacting light bosons are expected to be long-lived and leave very displaced vertex signals at experiments.
Among possible experiments, beam dump experiments are especially suitable for searching such bosons.
The bosons can be created from incoming beam on a target, travel long distance, and decay to electron-positron pairs near the detector distant from the target.
In the 1980s, several beam dump experiments were carried out to search for light neutral scalar bosons, such as axions, by using proton beam \cite{Bergsma:1985qz} and electron beam \cite{Konaka:1986cb,Riordan:1987aw,Bjorken:1988as,Bross:1989mp,Davier:1989wz}, and their experimental constraints were later translated to those on gauge bosons in Refs. \cite{Gninenko:2012eq} and \cite{Bjorken:2009mm,Andreas:2012mt}, respectively.
At present, the results of the beam dump experiments have been applied to various types of light boson models, and comprehensive analyses including constraints from collider experiments have also been done, e.g., see Refs. \cite{Bauer:2018onh}, \cite{Fabbrichesi:2020wbt,Caputo:2021eaa}, and \cite{Winkler:2018qyg} for the $B-L$ and the $L_\alpha-L_\beta$ gauge boson ($\alpha,\beta=e,\mu,\tau$), the dark photon, and the dark higgs boson, respectively. 
Also, in Refs. \cite{Kanemura:2015cxa,Sakaki:2020mqb,Asai:2021xtg,Asai:2021ehn}, the prospect of a new beam dump experiment at the future ILC experiment is studied.

Thought, beam dump experiments are probably the best probe of searching for feebly-interacting 
light bosons, the constraints are derived by assuming the simplest model setup, that is, 
only a single particle and a single coupling constant are introduced.
In the literature, however, several extensions have been proposed recently.

In this paper, given the current situation, we attempt to enlarge the availability of beam dump experiments for the extended scenarios.
We are especially interested in extensions including Charged Lepton Flavor Violation (CLFV) \cite{Heeck:2016xkh,Altmannshofer:2016brv,Iguro:2020rby,Davoudiasl:2021haa}.
As is well known, CLFV is one of the most evident signals for new physics beyond the SM.
Many experimental searches have been carried out, and in most cases muonic processes like $\mu \rightarrow eee$ and $\mu \rightarrow e\gamma$ place the tightest upper bounds on CLFV couplings.
On the other hand, in the sub-GeV mass regions, electron beam dump experiments can possibly exclude parameter regions below the upper bounds from the muonic processes, and new bounds could be obtained.
To illustrate the exclusion regions of electron beam dump experiments, we consider three benchmark scenarios for a new light boson, i.e., a scalar boson having Yukawa interactions, a gauge boson having vectorial interactions, and a gauge boson having dipole interactions.
For these, we introduce CLFV couplings in the electron ($e$) and muon ($\mu$) sector as well as Charged Lepton Flavor Conserving (CLFC) couplings, and derive the constraints from the E137 experiment, which usually provides the strongest constraints.

This paper is organized as follows.
In Sec.~\ref{sec:models}, we introduce three types of CLFV interactions analyzed in this work.
In Sec.~\ref{sec:production}, the production cross sections of light bosons through bremssthralung processes and the formula to calculate the number of signal events are given.
In Sec.~\ref{sec:constraints}, the constraints on the light bosons with CLFV coupling by 
the E137 experiment are derived, and the results are compared with the existing bounds from 
the muon CLFV decays.
In Sec.~\ref{sec:lfv-decay-signal}, the impact of searches for flavor violating decays are discussed.  
Section \ref{sec:summary} is devoted to summary.

\section{Interaction Lagrangian}
\label{sec:models}
We start our discussion by introducing interaction Lagrangians analyzed in this work.
We consider three types of interactions, i.e., scalar-, vector- and dipole-type interactions, 
and briefly discuss the origin of these interactions in mind a two Higgs doublet model, a gauged $\lmlt$ model  \cite{Foot:1990mn,He:1990pn,Foot:1994vd} and a dark photon model, respectively.
In the following, we restrict our analyses only to CLFC and CLFV in the $e\mu$ sector.

\subsection{Scalar interactions in two Higgs doublet model}
Firstly, we give the scalar-type interaction Lagrangian. 
Such an interaction can be obtained in models with an extra leptophilic Higgs doublet scalar.
The new Higgs doublet scalar is assumed to couple to only leptons and to contribute to mass generation of the charged leptons.
In this case, the mass matrix of the charged leptons consists of two parts originating from each Higgs doublet after developing vacuum expectation values (VEVs).
In general, those two mass matrices are not necessarily diagonalized simultaneously, and the misalignment of the mass matrices generates CLFV interactions after the diagonalization. 
To avoid large contributions from the CLFV interaction by the SM Higgs boson, 
we assume the VEV of the extra Higgs doublet field to be much smaller than the electroweak scale. Then, the charged lepton masses 
are mostly determined by mass matrix with the VEV of the SM Higgs boson.
In such a case, the CLFV interactions are given from the Yukawa terms of the extra scalar boson.

The relevant interaction Lagrangian is given by
\begin{align}
\mathcal{L}_{\rm scalar} = 
\sum_{\ell=e,\mu, \tau} y_\ell \overline{\ell_L} \phi \ell_R
+y'_{e\mu}\overline{e_L} \phi \mu_R
+y'_{\mu e}\overline{\mu_L} \phi e_R
+h.c.~,
\label{eq:Lscl}
\end{align}
where $\phi$ is the extra scalar boson, and $y_\ell$ and $y'_{e\mu (\mu e)}$ are 
the CLFC and CLFV coupling constants, respectively.
Here we omit the interactions obtained from the mixing between $\phi$ and the SM Higgs boson, due to the assumption of a small VEV for $\phi$.

From \eqref{eq:Lscl}, the total decay width of $\phi$ is given by 
\begin{align}
    \Gamma_{\mathrm{tot}} = 
      \sum_{\ell=e,\mu,\tau}\Gamma(\phi \rightarrow \ell\bar{\ell})
      + \Gamma(\phi \rightarrow e\bar{\mu})
      + \Gamma(\phi \rightarrow \mu\bar{e})~.
\end{align}
Here the partial decay widths into the charged leptons are given by \cite{DiazCruz:1999xe,Arganda:2004bz}
\begin{align}
\Gamma(\phi \rightarrow \ell\bar{\ell'})
= \frac{S^2}{8\pi}m_\phi
\left[ 1-\frac{(m_\ell+m_{\ell'})^2}{m_\phi^2} \right]^{3/2}
\sqrt{ 1-\frac{(m_\ell-m_{\ell'})^2}{m_\phi^2} }~,
\end{align}
where $S = y_\ell$ or $y'_{e\mu (\mu e)}$ for the CLFC- and CLFV-decays, respectively.

\subsection{Vector interactions in $L_\mu - L_\tau$ model}
Next, we give the vector-type interaction Lagrangian. 
The interactions of the charged leptons with a new vector boson can be 
obtained by extending the gauge sector of the SM. For minimality, we consider 
an extra U(1) gauge symmetry in this work. 
CLFV vector interactions can be generated at both tree and loop levels. 
At tree level, the following conditions should be satisfied: 
1) the different gauge charge assignments among charged lepton flavors, and 
2) the misalignment between the interaction and mass eigenstates of the charged leptons. Then, CLFV interactions appear in the gauge sector after diagonalizing the mass matrix of the charged leptons.
It should be noticed that the above conditions imply that the extra 
U(1) symmetry must be flavor-dependent. Therefore flavor universal gauge symmetries, such as the U(1)$_{B-L}$ gauge symmetry, can not induce the CLFV interactions in this way.
One of the viable examples is the U(1)$_{\lmlt}$ gauge symmetry with extra Higgs doublets.
Under the U(1)$_{\lmlt}$ gauge symmetry, the lepton flavors are differently charged: only 
the mu and the tau leptons ($\tau$) are charged as $+1$ and $-1$, respectively.
Moreover, if there exist extra Higgs doublets charged under the U(1)$_{\lmlt}$ gauge symmetry, off-diagonal elements in the charged lepton mass matrix are induced \cite{Foot:1994vd}.
At loop level, CLFV vector interactions will be induced through the CLFV scalar 
loop for massive gauge bosons. 
In this case, it is unnecessary for the gauge symmetry to be 
flavor-dependent. 
In this work, we only study the CLFV vector interactions at the tree level. However, analyses for the loop-induced CLFV interactions are essentially the same and will be translated by replacing the couplings with loop-induced ones in our results.

For general discussions, we parametrize the CLFV coupling by $\theta$ which is the 
mixing angle between electron and muon.
The mass and flavor eigenstates are 
connected by this mixing angle. 
Then, the Lagrangian of the vector interaction in mass eigenstate is given by
\begin{align}
\mathcal{L}_{\mathrm{vector}} &= g' Z'_\rho (
s^2~ \overline{e} \gamma^\rho e 
+ c^2~ \overline{\mu} \gamma^\rho \mu
+ sc~ \overline{\mu} \gamma^\rho e 
+ sc~ \overline{e} \gamma^\rho \mu) \nonumber \\
& \qquad +g' Z'_\rho (- \overline{\tau} \gamma^\rho \tau
+ \overline{\nu_\mu} \gamma^\rho \nu_\mu
- \overline{\nu_\tau} \gamma^\rho \nu_\tau
)~,
\label{eq:Lvec}
\end{align}
where $Z'$ is the gauge boson and $g'$ is the gauge coupling constant of the U(1)$_{\lmlt}$ gauge symmetry, 
and $s=\sin\theta$ and $c=\cos\theta$, respectively. Here $\nu_\mu$ and $\nu_\tau$ are left-handed 
muon and tau neutrinos. 
It should be noted that, in general, there also can exist interactions through the kinetic mixing. Such interactions conserve lepton flavor and are independent of the mixing angle. 
Then, the flavor conserving productions and decays of $Z'$ are 
modified. Although analyses of such a situation will lead more general constraints, the increase of the parameters will make the analyses  complicated.
For simplicity, we assume that contributions from the kinetic mixing can be negligible, and omit the kinetic mixing throughout this paper.

Given the Lagrangian in Eq. (\ref{eq:Lvec}), the total decay width of $Z'$ is obtained as
\begin{align}
\Gamma_{\mathrm{total}} = 
\Gamma(Z' \rightarrow \nu\bar{\nu}) 
+ \sum_{\ell=e, \mu,\tau}\Gamma(Z' \rightarrow \ell\bar{\ell})
+ \Gamma(Z' \rightarrow e\bar{\mu})
+ \Gamma(Z' \rightarrow \mu\bar{e})~,
\end{align}
where the partial decay width into the neutrinos is given by
\begin{align}
\Gamma(Z' \rightarrow \nu\bar{\nu}) = \frac{g'^2}{12\pi}m_{Z'}~,
\end{align}
in the limit of massless neutrinos\footnote{Here we assumed neutrinos are Dirac particle. For Majorana neutrinos, the partial decay width is multiplied by $1/2$.}, while those into the charged leptons are
\begin{align}
\Gamma(Z' \rightarrow \ell\bar{\ell'}) 
&= \frac{V^2}{24 \pi} m_{Z'}~\lambda\left(
\frac{m_\ell^2}{m_{Z'}^2}, \frac{m_{\ell'}^2}{m_{Z'}^2}\right) \nonumber \\
&\quad \times \left[
2
- \frac{m_\ell^2 - 6 m_\ell m_{\ell'} + m_{\ell'}^2}{m_{Z'}^2}
- \frac{(m_\ell^2 - m_{\ell'}^2)^2}{m_{Z'}^4}
\right]~,
\end{align}
where $V=g's^2~(g' c^2)$ or $g'sc$ for the decays into $ee~(\mu\mu)$ or $e\bar{\mu}$ and $\bar{e}\mu$, respectively, and $\lambda(a,b)$ is the Kallen function defined as follows~:
\begin{align}
\lambda(a,b) = \sqrt{1 + a^2 + b^2 - 2a - 2b - 2 ab}~.
\label{eq:kallen}
\end{align}

\subsection{Dipole CLFV in Dark Photon Model}
Lastly, we give the dipole-type interaction Lagrangian.
We again consider an extra U(1) gauge symmetry and assume that there are no interactions between the new gauge boson $A'$ and the SM particles at tree level, similar to the dark photon model with vanishing kinetic mixing.
Even in this case, CLFV interactions, as well as CLFC interactions, can be induced at the loop level. 
For instance, suppose new complex scalar bosons charged under the extra symmetry exist and couple 
to the SM charged leptons with new fermions. The interaction between $A'$ and charged leptons 
can be generated at one-loop in which the new scalar bosons and fermions propagate. 
Here we consider the following dipole-type interactions:
\begin{align}
    \mathcal{L}_{\mathrm{dipole}} &=
    \frac{1}{2} \sum_{\ell=e,\mu,\tau} \mu_\ell \overline{\ell} \sigma^{\rho \sigma} \ell A'_{\rho \sigma}
    + \frac{\mu'}{2} 
      \left( 
          \overline{\mu} \sigma^{\rho \sigma}e 
        + \overline{e} \sigma^{\rho \sigma}\mu 
      \right) A'_{\rho \sigma}~,
\label{eq:Ldpl}
\end{align}
where $\mu_\ell$ and $\mu^\prime$ are CLFC and CLFV dipole couplings, respectively, and $A_{\rho\sigma}^\prime$ stands for the field strength of $A^\prime$. We assume that 
the dipole couplings are real.
One may imagine that there should exist similar CLFV interactions in which external $A'$ is replaced by photon, 
which are strictly constrained by the MEG~\cite{MEG:2016leq} and  BaBar~\cite{BaBar:2009hkt} experiments. 
However, it will be possible to suppress such dangerous electromagnetic dipole operators when the new gauge boson has an interaction vertex with neutral CP-even and odd scalars since the same vertex does not exist for the photon. 
One of such examples is the so-called dark photon model with dark Higgs particles 
\cite{Nomura:2020azp}.

Given the Lagrangian in Eq. (\ref{eq:Ldpl}), the total decay width of $A'$ is given by
\begin{align}
\Gamma_{\mathrm{total}} = 
\sum_{\ell=e,\mu,\tau}\Gamma(A' \rightarrow \ell\bar{\ell})
+ \Gamma(A' \rightarrow e\bar{\mu})
+ \Gamma(A' \rightarrow \mu\bar{e})~,
\end{align}
where
\begin{align}
\Gamma(A' \rightarrow \ell\bar{\ell'}) 
&= \frac{D^2}{12 \pi} m_{A'}^3 ~\lambda\left(\frac{m_\ell^2}{m_{A'}^2}, \frac{m_{\ell'}^2}{m_{A'}^2}\right) \nonumber \\
&\quad \times 
\left[
\frac{1}{2} 
+ \frac{1}{2}\frac{m_\ell^2 + 6m_\ell m_{\ell'} + m_{\ell'}^2}{m_{A'}^2}
- \frac{(m_\ell^2 - m_{\ell'}^2)^2}{m_{A'}^4}
\right],
\end{align}
$D = \mu_\ell$ or $\mu'$ for CLFC or CLFV decays, respectively, and $\lambda(a,b)$ is given in Eq. (\ref{eq:kallen}).

\section{Scalar and Gauge Boson Production in Beam Dump Experiments}
\label{sec:production}
\begin{figure}[t]
\centering
\includegraphics[width=0.4\textwidth]{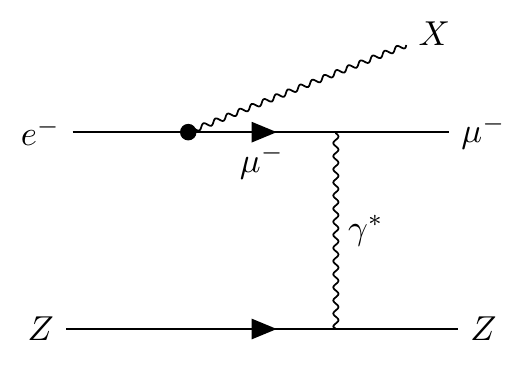} \quad
\includegraphics[width=0.4\textwidth]{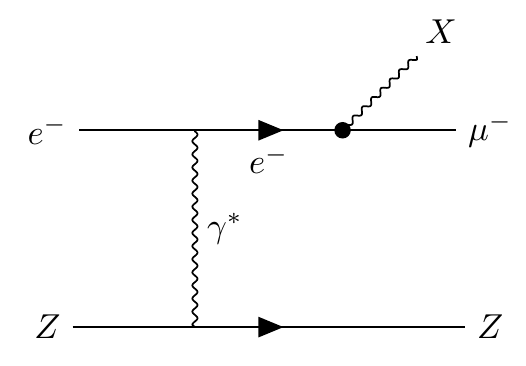}
\caption{
Light boson production through bremsstrahlung process with CLFV interaction by a target with the atomic number $Z$.
}
\label{fig:bremsstrahling-LFV}
\end{figure}

In electron beam dump experiments, a new light boson $X~(=\phi, Z', A')$ is produced by the bremsstrahlung process with nucleons in target materials.
When the light bosons have CLFV interactions with electron and muon, they also can be produced through 
flavor violating bremsstrahlung processes shown in Fig.~\ref{fig:bremsstrahling-LFV}.
The bremsstrahlung production process can be evaluated by using the cross section of electron-photon scattering, 
$e^- + \gamma \to \ell + X~(\ell=e,\mu)$, in the Weizs\"{a}cker-Williams approximation~\cite{vonWeizsacker:1934nji,Williams1935,Kim:1973he}.
In this section, we show the differential cross sections of the scalar and vector boson production process for 
the interactions given in the previous section. 
Then, we give formulae to calculate the number of events in electron beam dump experiments.

\subsection{Differential cross section of bremsstrahlung process}

With the improved Weizs\"{a}cker-Williams approximation, the differential cross section of bremsstrahlung process of the light boson production by a target with atomic number $Z$, $e^- + Z \to \ell^- + Z + X$ $(\ell = e, \mu,~X = \phi, Z', A')$, is calculated by that of scattering one, $e^- + \gamma \to \ell^- + X$, as follows~:
\begin{align}
    \frac{d \sigma_{\rm brems}}{d x} = \frac{\alpha \xi}{\pi} \frac{E_0 x \beta_X}{1-x} \frac{d \sigma_{\rm scat}}{d x}~,
\end{align}
where $\alpha$ is the electromagnetic fine structure constant, $E_0$ is the energy of the injected electron beam, $\beta_X = \sqrt{1-m_X^2/E_e^2}$ is the kinematical factor,  and $x = E_X / E_e$ with $E_{e (X)}$ being the energy of incident electron (produced light boson). 
The effective photon flux is denoted as $\xi$~\cite{Bjorken:2009mm}, and the definition of $\xi$ is given in Appendix~\ref{apdx:eff-photon-flux}.
The differential cross section of the scattering process 
with respect to $x$ is given by 
\begin{align}
\frac{d\sigma_{\rm scat}}{d x} = \frac{\alpha g^2_X}{2 E_0} \frac{1-x}{x} 
    &\left[ f_1(x) \frac{\tilde{U}_1^\ell}{m_{X}^2} + f_2(x) \frac{\tilde{U}_2^\ell}{E_e^2 x} \right. \nonumber \\
    &\quad \left. - f_3(x) 
    \bigg( x \frac{m_{X}^2 \tilde{U}_3^\ell}{(E_e^2 x)^2} 
    - ( 1 - a_1 x + r_e x^2) \frac{m_{X}^4 \tilde{U}_4^\ell}{(E_e^2 x)^3} \bigg)
    \right],
    \label{eq:diff-cross-sec}
\end{align}
where $g_X$ and $m_X$ are the coupling constant and mass of the light boson $X$, respectively. 
In the square bracket, $\tilde{U}_n^\ell$ ($n=1$-$4$) are functions of the maximal angle $\theta_{\mathrm{max}}$ determined
by the angular acceptance of the detector and $\eta_\ell$ given by
\begin{align}
    \eta_\ell= \frac{m_X^2}{E_e^2}\frac{1-x}{x^2} + \frac{m_e^2}{E_e^2} + \frac{m_\ell^2 - m_e^2}{E_e^2 x}~, \label{eq:eta-l}
\end{align}
where $m_\ell$ is the mass of lepton $\ell$. 
The definition and approximate form of $\tilde{U}_n^\ell$ are given in Appendix~\ref{apdx:tilde-u}. 
The functions $f_1(x)$, $f_2(x)$ and $f_3(x)$ depend on the final state lepton $\ell$ and the types of the light boson $X$. 
For convenience, we define 
\begin{align}
r_\ell = \frac{m_\ell^2}{m_X^2}~,
\end{align}
and 
\begin{subequations}
\begin{align}
a_1 &= 1 + r_e - r_\ell~, \\
a_2 &= 1 - r_e - r_\ell~,\\ 
a_3 &= 2 + r_e + r_\ell - 2 \sqrt{r_e r_\ell}~, \\
a_4 &=  2 - r_e -r_e^2 - r_\ell -r_\ell^2 + 6 \sqrt{r_e r_\ell} + 2r_e r_\ell~.
\end{align}
\end{subequations}
Note that $\ell = e$ corresponds to the case of the CLFC interactions  while $\ell=\mu$ to the CLFV interactions.
Thus, the difference between the CLFC and CLFV interactions arises in the third term of $\eta_\ell$ as well as $r_\ell$ and $g_X$.

For the scalar interaction given in \eqref{eq:Lscl}, $X=\phi$ and $g_\phi = 1$. The functions $f_{1,2,3}(x)$ 
in \eqref{eq:diff-cross-sec} are 
\begin{gather}
    f_1(x) = 0~,~~~
    f_2(x) = S_1 \frac{x^2}{2}~,~~~
    f_3(x) = (a_2 S_1 - 4 \sqrt{r_e r_\ell} S_2)(1-x)~,
    \label{eq:diff-cs-scalar}
\end{gather}
where $S_1 = |y_{e \mu}|^2 + |y_{\mu e}|^2$ and $S_2 = \mathrm{Re}(y_{e \mu} y_{\mu e})$ for the CLFV process 
($\ell = \mu)$, while $S_1 = 2|y_e|^2$ and $S_2 = \mathrm{Re}(y_e^2)$ for the CLFC $(\ell = e)$, 
respectively.
For the vector interaction in \eqref{eq:Lvec}, $X=Z'$ and $g_{Z'}=g'sc~(g's^2)$ for the CLFV (CLFC) process. 
The functions in \eqref{eq:diff-cross-sec} are 
\begin{gather}
    f_1(x) = 0~,~~~
    f_2(x) = 4 - 4 x + a_3 x^2~,~~~
    f_3(x) = 2 a_4 (1-x)~.
    \label{eq:diff-cs-vector}
\end{gather}
In the dipole case, $X=A'$ and $g_{A'} = \mu' m_{A'}~(\mu_e m_{A'})$ for the CLFV (CFLC) process, 
and 
\begin{gather}
    f_1(x) = 4x~,~~~
    f_2(x) = x( x + 2(r_\ell - r_e)(x-2))~,~~~
    f_3(x) = 2 a_4 (1-x)~.
    \label{eq:diff-cs-dipole}
\end{gather}

\begin{figure}[t]
\centering
\includegraphics[width=0.48\textwidth]{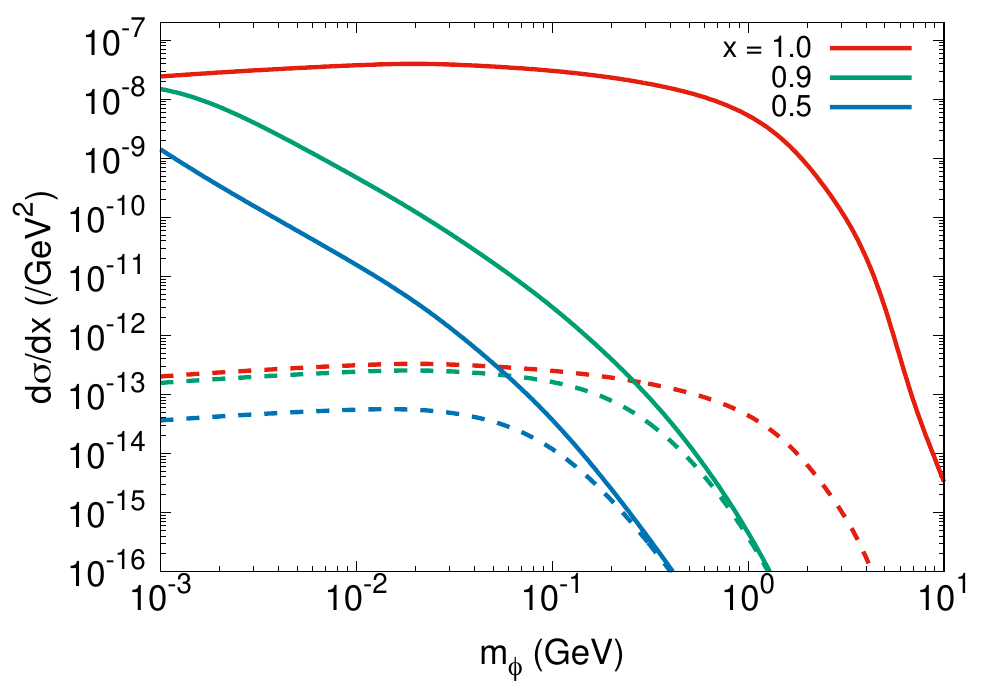} \quad
\includegraphics[width=0.48\textwidth]{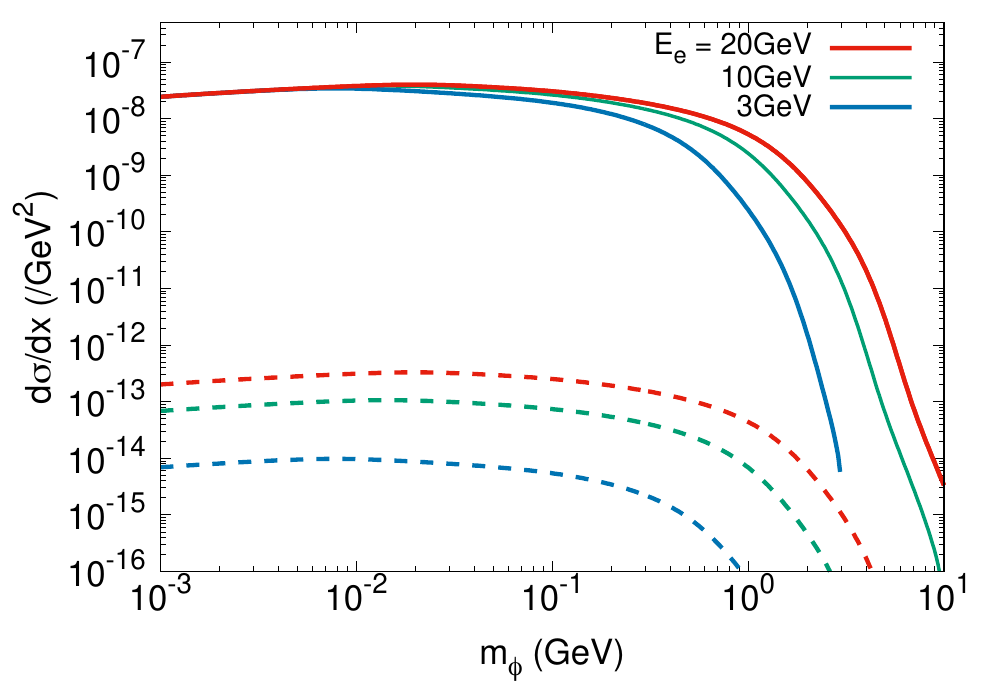}
\caption{
Differential cross section $d\sigma_{\mathrm{brems}}/dx$ for the scalar interaction.
Solid and dashed curves represent the CLFC and CLFV case, respectively, and the coupling constants are taken to be $y_e = y'_{e\mu}=y'_{\mu e} = 10^{-6}$.
Left panel: for $x=1.0$ (red), $0.9$ (green) and $0.5$ (blue), while fixing $E_e$ as 20 GeV.
Right panel: for $E_e=20$ (red), $10$ (green) and $3$ (blue) GeV, while fixing $x$ as 1.0.
}
\label{fig:diff-cross-sec-scalar}
\end{figure}
\begin{figure}[t]
\centering
\includegraphics[width=0.48\textwidth]{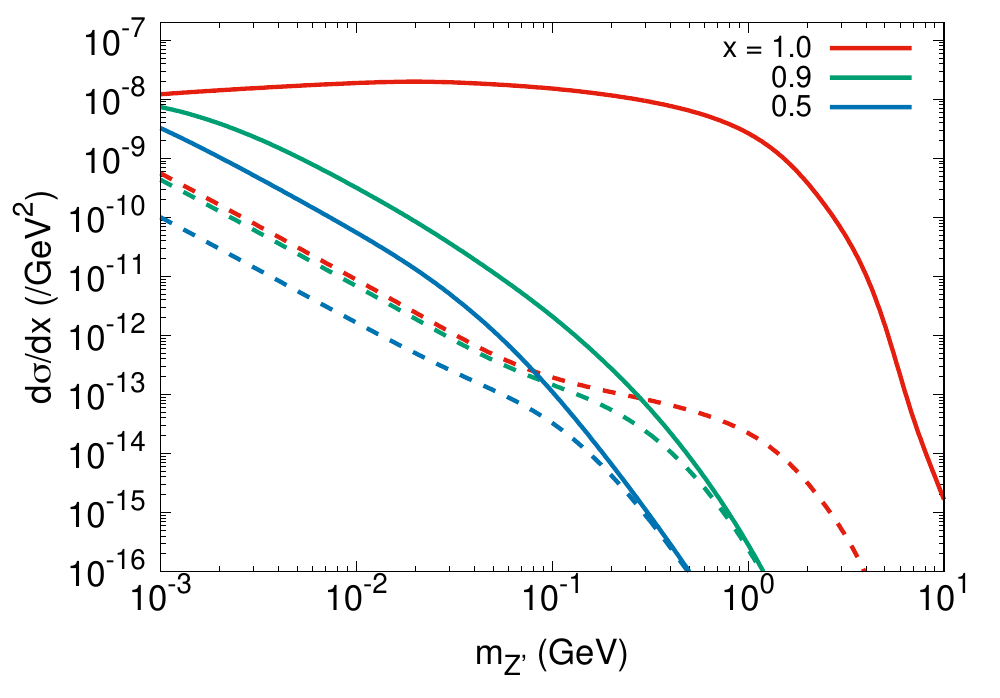} \quad
\includegraphics[width=0.48\textwidth]{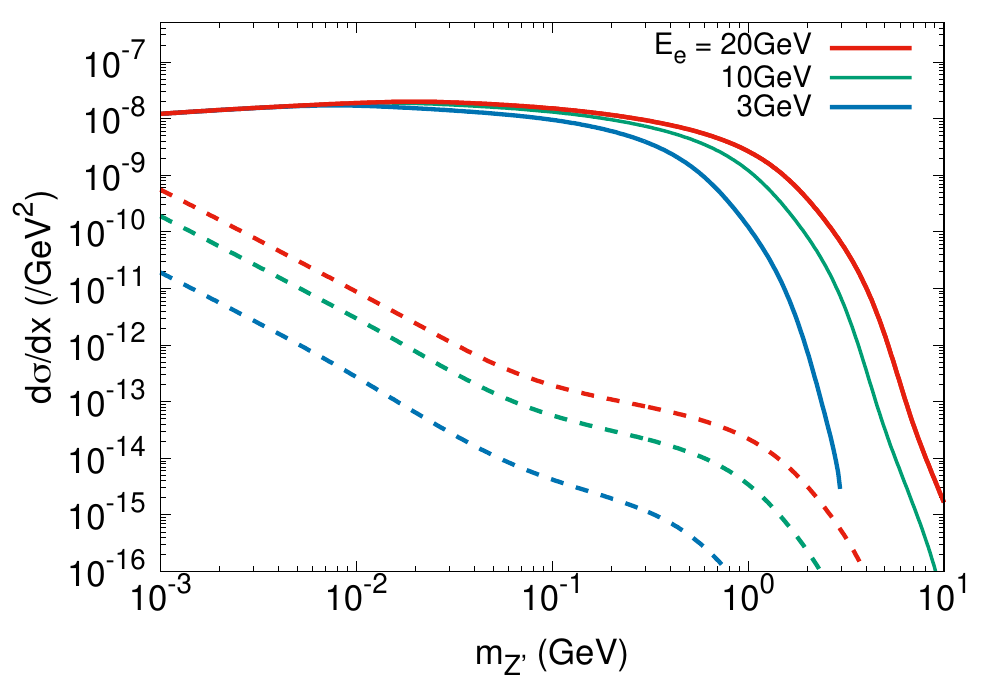}
\caption{
The same plots as Fig.~\ref{fig:diff-cross-sec-scalar} for the vector interaction. 
The coupling constants are taken to be $g'=10^{-6}$ and $\theta=\pi/4$.
}
\label{fig:diff-cross-sec-vector}
\end{figure}
\begin{figure}[t]
\centering
\includegraphics[width=0.48\textwidth]{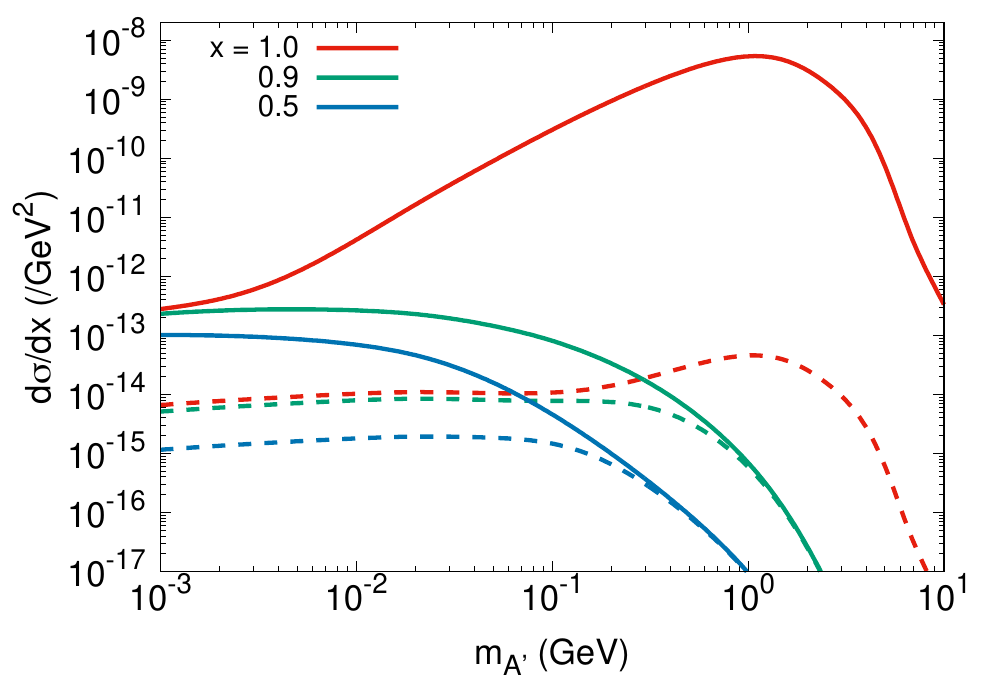} \quad
\includegraphics[width=0.48\textwidth]{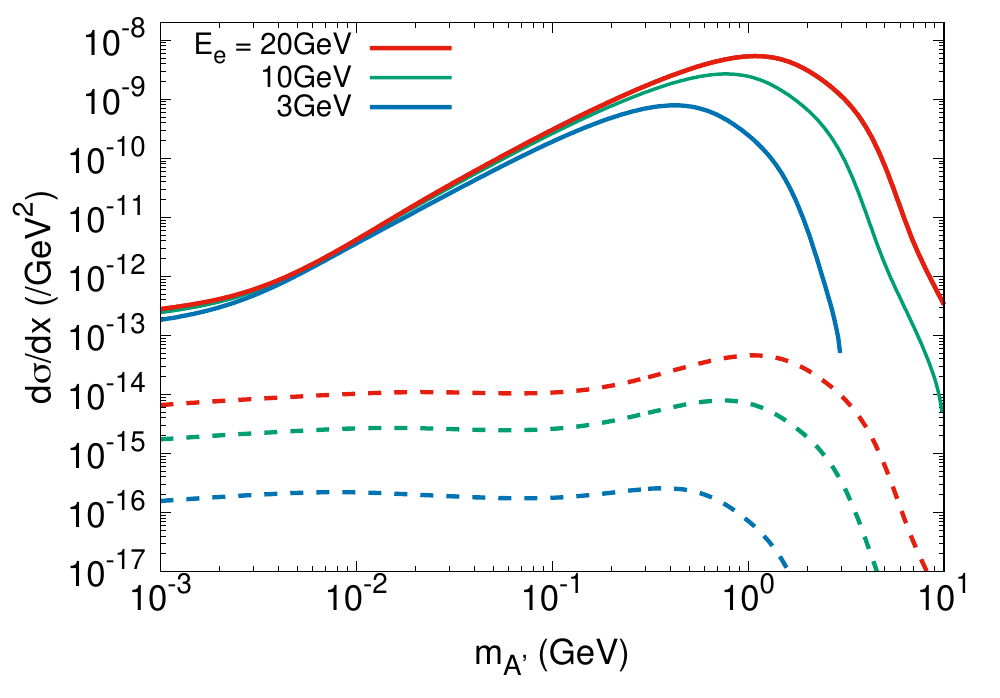}
\caption{
The same plot as Fig.~\ref{fig:diff-cross-sec-scalar} for the dipole interaction. 
The coupling constants are taken to be $\mu_e = \mu' = 10^{-6}$ GeV$^{-1}$ the 
for dipole interactions.
}
\label{fig:diff-cross-sec-dipole}
\end{figure}

Figures \ref{fig:diff-cross-sec-scalar}, \ref{fig:diff-cross-sec-vector}, and \ref{fig:diff-cross-sec-dipole} show the differential cross sections of the bremsstrahlung process for the scalar, vector, and dipole interactions with Eqs.~(\ref{eq:diff-cs-scalar}), (\ref{eq:diff-cs-vector}), and (\ref{eq:diff-cs-dipole}), respectively, as a function 
of the $X$ boson mass.
In the left panels, $E_e$ is fixed to be $20$ GeV, and $x$ is taken to be 
$1.0$ (red), $0.9$ (green), and $0.5$ (blue), respectively. In the right panels, 
$E_e$ is varied as $20$ (red), $10$ (green), and $3$ (blue) GeV, respectively, and 
$x$ is fixed to be $1$. Solid and dashed curves correspond to the CLFC and CLFV interactions, respectively.

In figure \ref{fig:diff-cross-sec-scalar}, the coupling constants are taken to be $y_e = y'_{e\mu}=y'_{\mu e} = 10^{-6}$ as an illustrating example. In the left panel, the CLFV differential cross sections are much 
smaller than the CLFC ones with the same $x$. This is because $\tilde{U}^\ell_n$ are decreasing functions of $\eta_\ell$, and $\eta_\mu$ is larger than $\eta_e$. 
The CLFC differential cross sections increase by order of magnitudes as $x$ approaches unity because 
the second term in \eqref{eq:diff-cross-sec} is much larger than other terms.
These with $x<1$ also rapidly decrease as $m_\phi$ becomes large 
since $\eta_e$ scales as $m_\phi^2/E_e^2$. On the other hand, the CLFV differential cross sections are rather constant because 
$\eta_\mu$ scales as $m_\mu^2/E_e^2$ for $m_\phi < m_\mu$.
For $m_\phi$ larger than $m_\mu$, $r_\mu \ll 1$ and $\eta_\mu \simeq \eta_e$, therefore  
the CLFC and CLFV cross sections asymptotically approach to each other. 
In the right panel, the CLFC differential cross sections with $x=1$ are similar to different $E_e$ while 
the CLFV ones change order of magnitudes. For $x=1$, \eqref{eq:diff-cross-sec} for the CLFC process is approximated as 
\begin{align}
    \frac{d \sigma_{\mathrm{brems}}}{dx}\simeq \frac{\alpha^2}{2 \pi}  \xi \beta_{\phi} 
     \frac{S_1}{4 m_e^2}~,
\end{align}
where we have used $\tilde{U}_2^e \simeq 1/2 \eta_e = E_e^2/2 m_e^2$. The effective photon flux is almost constant for $m_\phi < 1$ GeV. Thus, the electron energy dependence only comes from $\beta_\phi$ which is almost unity unless close to the threshold.
In the CLFV process, on the other hand,  the electron energy dependence remains in $\tilde{U}_2^e/E_e^2$.
From both panels, it can be understood that the main contributions to the signal event come from the CLFC process with $0.9 \lsim x \lsim 1$.

Figure \ref{fig:diff-cross-sec-vector} shows the same plots with Fig.~\ref{fig:diff-cross-sec-scalar} for the vector interaction. 
The coupling constant and the mixing angle are fixed to $g'=10^{-6}$ and $\theta=\pi/4$. 
Contrary to the scalar interaction case, the CLFV differential cross sections decrease as $m_{Z'}$ increases. This is because $f_2(x)$ in \eqref{eq:diff-cs-vector} is dominated by $r_\mu x^2$ for small $m_{Z'}$. 
The behavior of the CLFC differential cross sections can be understood in the same ways as in the scalar interaction case. 
Figure \ref{fig:diff-cross-sec-dipole} also shows the same plots for the dipole interaction. 
The dipole moments are fixed to $\mu_e = \mu' = 10^{-6}$ GeV$^{-1}$. 
In this case, the differential 
cross sections are proportional to $m_{A'}^2$. These increase as $m_{A'}$ increases, then goes to zero as 
the mass reaches to the kinematical threshold. 
From Figs.~\ref{fig:diff-cross-sec-vector} and \ref{fig:diff-cross-sec-dipole}, the dominant contributions to the signal events also come from the CLFC differential cross sections with $0.9 \lsim x \lsim 1$ for the vector and dipole 
interaction case, respectively.

\subsection{Number of signal events}
\label{subsec:signal}
The number of the signal events can be calculated by the following 
formula \cite{Bjorken:2009mm},
\begin{align}
    N&=N_e \frac{N_{\mathrm{avo}} X_0}{A} \sum_{\ell = e,\mu} \int_{m_X}^{E_0 - m_\ell} dE_{X} \int_{E_X+m_\ell}^{E_0} dE_e \int^{T_{\mathrm{sh}}}_0 dt \nonumber \\
    &\quad \times \left[
    I_e(E_0, E_e, t) \frac{1}{E_e} \left. \frac{d\sigma_{\rm brems}}{dx}\right|_{x = \frac{E_X}{E_e}} e^{-L_{\mathrm{sh}}/L_X} (1-e^{-L_{\mathrm{dec}}/L_X})
    \right] \mathrm{Br}(X \to e^+ e^-)~,
    \label{eq:Nsig}
\end{align}
where $N_e$ is the number of electrons in the injected beam, $N_{\rm avo} \simeq 6 \times 10^{23}~{\rm mol}^{-1}$ the Avogadro's number, $X_0$ the radiation length of the target, $A$ the target atomic mass 
in g/mol, and $T_{\rm sh} \equiv \rho_{\rm sh} L_{\rm sh} / X_0$ with $\rho_{\rm sh}$ being the density 
of the shield. The length of shield and decay region are denoted as $L_{\rm sh}$ and $L_{\rm dec}$, 
respectively. The decay length in laboratory frame and branching ratio of $X$ are denoted by $L_X$ and 
Br$(X \to e^+ e^-)$, respectively.
The energy distribution of electrons after passing through $t$ radiation lengths in the beam dump is 
denoted by $I_e$ and given by~\cite{Tsai:1986tx}
\begin{align}
    I_e(E_0, E_e, t) = \frac{1}{E_0} \frac{\left[ \ln (E_0 / E_e) \right]^{b t - 1}}{ \Gamma(b t)}~,
\end{align}
where $b = 4/3$~. This energy distribution function sharply peaks around $t \simeq 0$ for $E_e = E_0$, 
and the $t$ integration can diverge. To avoid such divergence, we split the $t$ integration into two 
parts by a cut $t_\mathrm{cut}$, following Ref.~\cite{Andreas:2013xxa}.

\section{Electron Beam Dump Constraints}
\label{sec:constraints}
In this section, we show the constraints from electron beam dump experiments, E137 \cite{Bjorken:1988as}, 
for the scalar, vector, and dipole interactions with CLFV. 
%
\begin{table}[t]
\begin{tabular}{|cccccc|} \hline
target & \begin{tabular}{c} radiation length \\ $X_0$ [g/cm$^2$] \end{tabular} & \begin{tabular}{c} beam energy \\ $E_0$ [GeV] \end{tabular} &   \begin{tabular}{c} total electrons \\ $N_e$ \end{tabular} &  \begin{tabular}{c} shield length \\ $L_\mathrm{sh}$ [m] \end{tabular} &  \begin{tabular}{c}decay volume length \\ $L_\mathrm{dec}$ [m] \end{tabular} \\ \hline
\ce{^{26.98}_{13}Al} & 24.01 & 20 & $1.86 \times 10^{20}$ & 179 & 204 \\ \hline
\end{tabular}
\caption{Setup of the E137 experiment.}
\label{tab:E137setup}
\end{table}
%
In the E137 experiment, the electron beam with energy $20$ GeV is injected into an aluminium target.
The total electrons dumped in the experiment is $1.86 \times 10^{20}$. The detector consists of 
electromagnetic shower counter with dimensions $2$ m $\times$ $3$ m in the first phase, and 
$3$ m $\times$ $3$ m in the second phase. It has $204$ m decay region placed at $179$ m downstream to
the dump\footnote{
One may consider the produced muon could penetrate the shield and be observed by the detector in beam dump experiments. 
Using the continuous-slowing-down-approximation range \cite{Zyla:2020zbs}, the range for the muon to be stopped is estimated at $\sim 50$m at most for the E137 experiment. The range is $1/3$ times shorter than the shield length. Therefore, the produced muon would be stopped in the shield.}. 
We summarize the setup of the E137 experiment in Tab.~\ref{tab:E137setup}.
In order to take into account the geometrical acceptance, we set $\theta_{\rm max}$ in Eq. (\ref{eq:tilde-U}) as $\theta_{\rm max} = 0.00392$ rad.
The E137 collaboration reported null results for the energy above $3$ GeV in the search for axion-like particles, so that we regard the regions predicting $N > 3$ events as the 95\% C.L. exclusion regions and set the lower cut of $3$ GeV on the $E_X$ integral in Eq. (\ref{eq:Nsig}). 
%
\begin{figure}[t]
\centering
\includegraphics[width=0.6\textwidth]{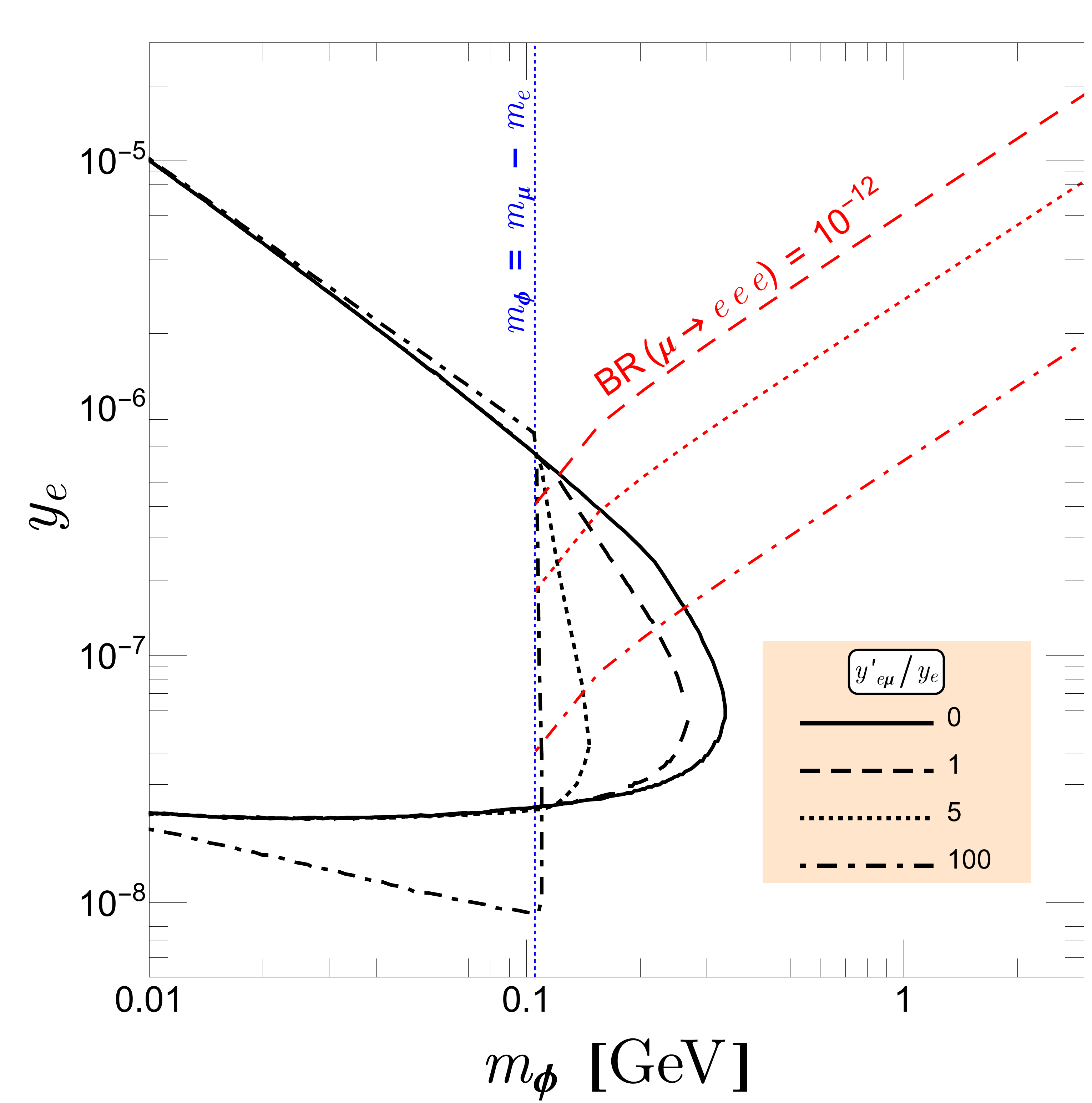}
\caption{
Exclusion region from the E137 experiment for the scalar interaction. The CLFV coupling is taken to 
$y'_{e\mu}/y_e =0$ (solid), $1$ (dashed), $5$ (dotted) and $100$ (dot-dashed). 
Red curves represent the constraints from $\mu \to eee$, and blue vertical line corresponds to $m_\phi = m_\mu - m_e$.
}
\label{fig:scalar-constraint}
\end{figure}

%
Figure \ref{fig:scalar-constraint} shows the exclusion regions in $m_\phi$-$y_e$ plane at $95$\% C.L. from the E137 for the scalar case. 
For simplicity, we assumed $y'_{e\mu} = y'_{\mu e}$ and $y_e,~y_{e\mu}$ are real.
Solid, dashed, dotted and dot-dashed curves correspond to $y'_{e\mu}/y_e =0,~1,~5$ and $100$, respectively. 
Among them, $y_{e\mu}'/y_e = 0$ (the solid curve) corresponds to the case of no CLFV coupling, and we find that the E137 excludes large parameter regions below $m_\phi = 0.35$ GeV.
For the cases of nonzero $y'_{e\mu}$, there are strong constraints from CLFV decays of the muon.
The regions above the red curves are excluded by $\mu \to eee$, while most regions of the left side of the blue vertical line are excluded by $\mu \to e \phi$.
The upper bounds of these decay branching ratios are listed in Table \ref{tab:constraints}, and the partial widths of these CLFV decays 
are given in Appendix~\ref{apdx:clfv-deca-width}.
Note that constraints from $\mu \rightarrow e\gamma$ are weaker than those from $\mu \rightarrow eee$ since the decay width of $\mu \rightarrow e\gamma$ is suppressed by the electromagnetic coupling and a loop factor in comparison with that of $\mu \rightarrow eee$.
As can be seen from the figure, most of the parameter regions are excluded by these constraints.
Nevertheless, we find that our 95\% C.L. limit further excludes the parameter regions unconstrained by $\mu \to eee$ and $\mu \to e \phi$.
%
\begin{table}[t]
\begin{tabular}{|c|c|c|c|} \hline
Br($\mu \to eee$) & \multicolumn{3}{|c|}{Br($\mu \to e X$)} \\  \hline
~~~$m_X > m_\mu - m_e$~~~ & ~~~$13 < m_X < 80$ ~~~ & 
~~~$47.8 < m_X < 95.1$ ~~~ & 
~~~$98.1 < m_X < 103.5$ ~~~ \\ \hline
$ < 1.0 \times 10^{-12}$ \cite{SINDRUM:1987nra} & $ \lsim 5.8 \times 10^{-5}$ \cite{TWIST:2014ymv} & 
$ \lsim 10^{-5}$ \cite{PIENU:2020loi} & $ \lsim 2 \times 10^{-4}$ \cite{Derenzo:1969za} \\ \hline
\end{tabular}
\caption{Upper bounds on the branching ratios of $\mu \to eee$ and $\mu \to e X$. Masses are in MeV.}
\label{tab:constraints}
\end{table}
%
\begin{figure}[t]
\centering
\includegraphics[width=0.6\textwidth]{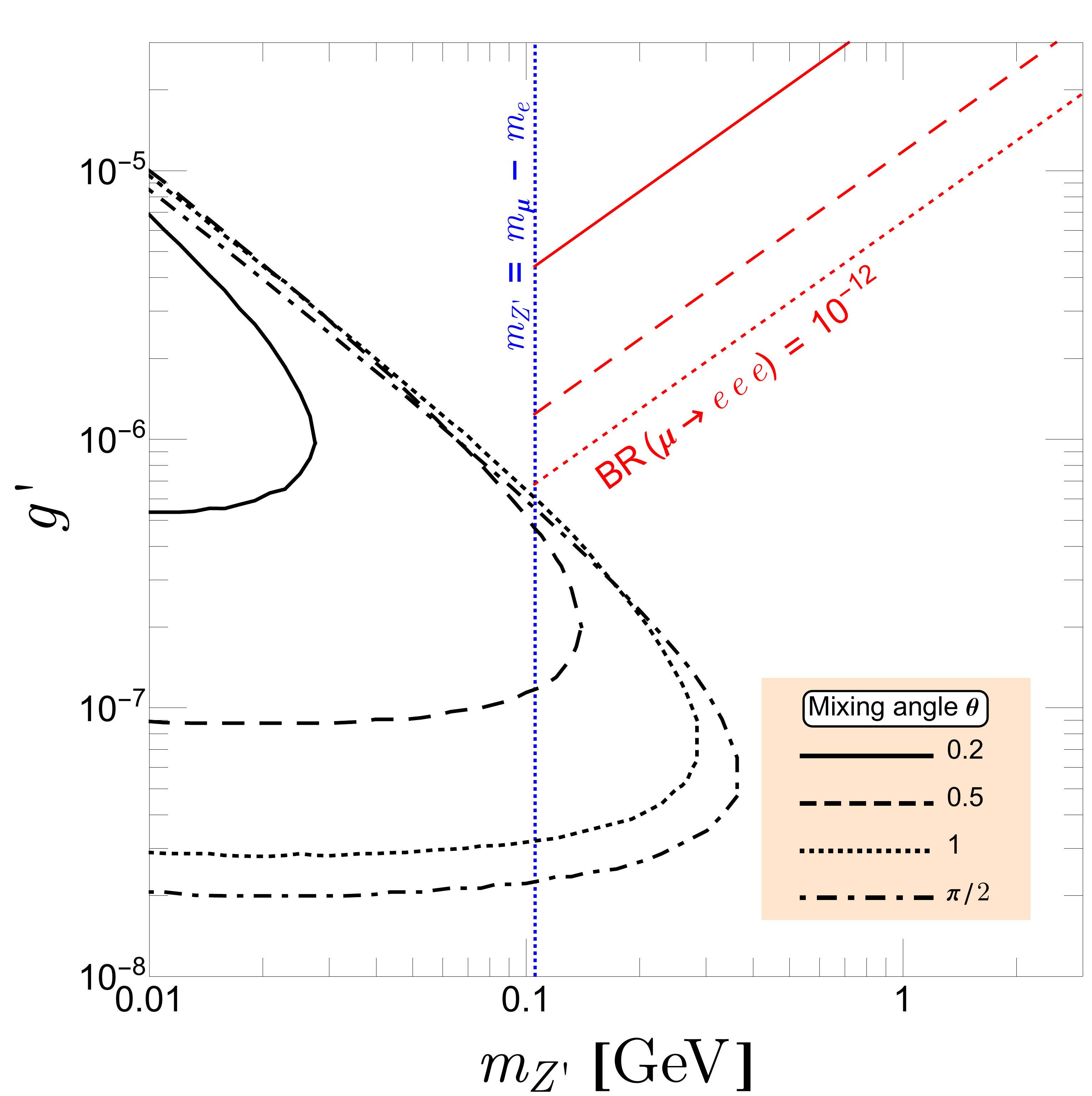}
\caption{
The same plots as Fig.~\ref{fig:scalar-constraint} for the vector interaction. The mixing angle is taken to 
$0.2$ (solid), $0.5$ (dashed), $1$ (dotted) 
and $\pi/2$ (dot-dashed), respectively.
}
\label{fig:vector-constraint}
\end{figure}
%
The exclusion region shifts to lighter $m_\phi$ as $y'_{e\mu}/y_e$ increases. 
As we explained in Sec.~\ref{sec:production}, the differential cross sections are mainly determined 
by the CLFC ones for $y'_{e\mu} < 10$. 
Therefore, in this case, the effects of the flavor violating couplings appear only in the total decay width.
Once the threshold of $\phi \to e \mu$ opens, the decay length and branching ratio of $\phi \to e e$ become smaller. 
Then, the expected number of the signal events reduces, resulting in the exclusion regions shown in Figure \ref{fig:scalar-constraint}.
For $y'_{e\mu} = 100$, 
contributions from the CLFV differential cross sections are not negligible, making the exclusion regions wider especially in the small $y_e$ regions.
Such a large $y_{e\mu}^\prime$, however, makes the decay length and the branching ratio so small that no signal events are expected above the threshold of $\mu\rightarrow e\phi$.

\begin{figure}[t]
\centering
\includegraphics[width=0.6\textwidth]{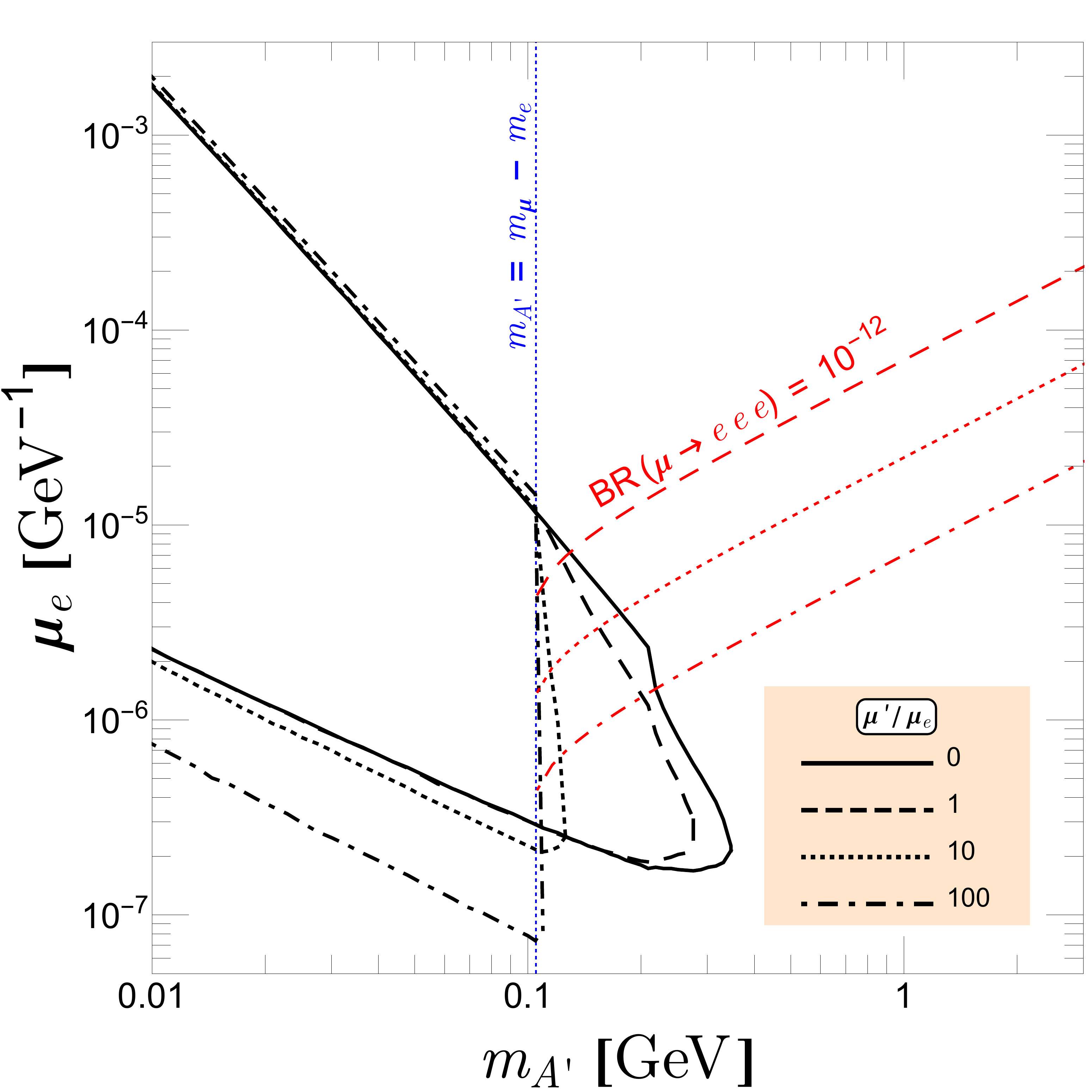}
\caption{
The same plots as Fig.~\ref{fig:scalar-constraint} for the dipole interaction. The dipoles are taken to 
$\mu'/\mu_e = 0$ (solid), $1$ (dashed), $10$ (dotted) 
and $100$ (dot-dashed), respectively.
}
\label{fig:dipole-constraint}
\end{figure}
%
Figure \ref{fig:vector-constraint} shows the exclusion plot in $m_{Z'}$-$g'$ plane for the vector interaction. The mixing angle is varied as $\theta = 0.2$ (solid), $0.5$ (dashed), $1$ (dotted) 
and $\pi/2$ (dot-dashed), respectively. 
The red and blue curves represent the same constraints as in Fig.~\ref{fig:scalar-constraint}.
In contrast to the case of the scalar interaction, the exclusion regions become wider as the CLFV mixing becomes large.
This is because a nonzero $\theta$ induces interactions between $Z'$ and electrons\footnote{
Even in the case of $\theta=0$, interactions between $Z'$ and electrons arise at one loop level, which yields narrow exclusion regions around $m_{Z'} = 0.001\,\mathchar`-\,0.003$ GeV \cite{Bauer:2018onh}.
}, 
increasing the production cross sections and the branching ratio of $Z' \rightarrow ee$. 
For $\theta > 0.5$, the constraint from E137 experiment excludes the parameter regions for $m_{Z'} > m_\mu + m_e$, which are not constrained by $\mu \to eee$ and $\mu \to e Z'$. 
When $\theta=\pi/2$, the interaction to muon vanishes and the exclusion region coincides with 
that for the minimal U(1)$_{L_e - L_\tau}$ model \cite{Bauer:2018onh}.

Figure \ref{fig:dipole-constraint} is the same plot as Fig.~\ref{fig:scalar-constraint} and \ref{fig:vector-constraint} for the dipole interaction in $m_{A'}$-$\mu_e$ plane. 
The CLFV dipole is taken to $\mu'/\mu_e = 0$ (solid), $1$ (dashed), $10$ (dotted) and $100$ (dot-dashed), respectively. In this case, the constraint from the E137 experiment also excludes the parameter region 
below the $\mu \to eee$ limit. The behavior of the exclusion region is similar to the scalar 
interaction case.

\section{Flavor violating decay signal}
\label{sec:lfv-decay-signal}
In the previous section, we have only considered the CLFC decay of $X \to ee$ as the signal of the 
vector and scalar boson, based on the setup and analyses of the E137 experiment. Under the presence of the CLFV interactions, the light bosons can decay into $e\bar{\mu}$ and $\bar{e}\mu$ above its kinematical threshold. This decay mode is a smoking gun signature of CLFV interactions in the dark sector. 
Searches for these decays will bring further information on the CLFV couplings. 

%
\begin{figure}[t]
\centering
\includegraphics[width=0.6\textwidth]{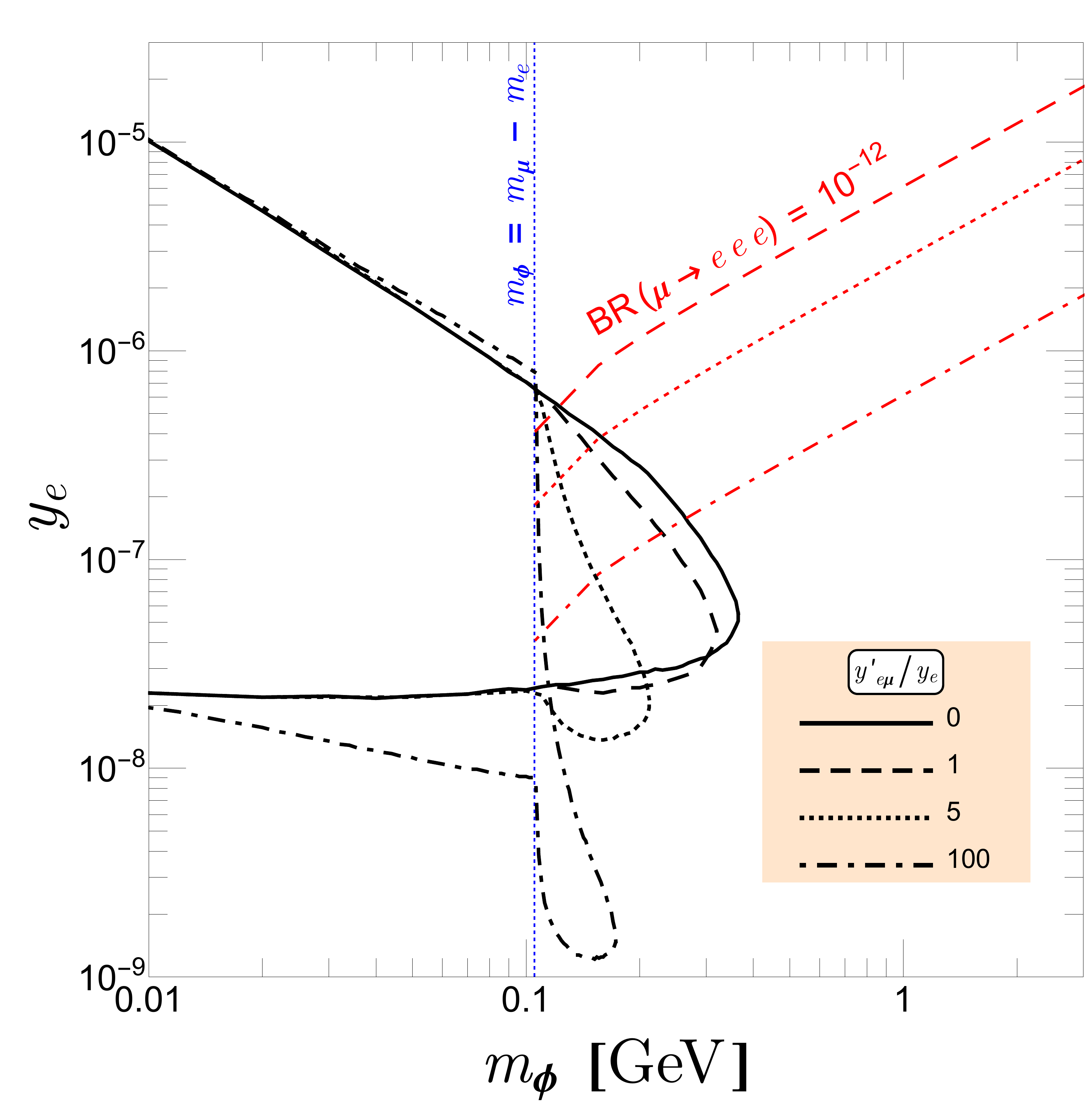}
\caption{
Presumed exclusion region to $\phi \to ee,~e\mu,~\mu\mu$ decays in the E137 experiment for the scalar interaction. The parameters are taken as $y'_{e\mu}/y_e = 0$ (solid), $1$ (dashed), $5$ (dotted) and $100$ (dot-dashed).
}
\label{fig:clfv-constraint}
\end{figure}
To illustrate the impact of the CLFV decay searches, we have demonstrated analyses for the scalar CLFV 
decays with the setup of the E137 experiment in Table.~\ref{tab:E137setup}.
Figure \ref{fig:clfv-constraint} shows presumed excluded regions with $95$\% C.L. 
for $\phi \to ee,~e\mu,~\mu\mu$.
The solid, dashed, dotted and dot-dashed curves correspond to $y'_{e\mu}/y_e =0$, $1$, $5$ and $100$, 
respectively.
Compared with Fig.~\ref{fig:scalar-constraint}, smaller coupling and larger mass regions can be 
excluded above the $e\mu$ threshold. 
This is because, for $m_\phi \geq m_e + m_\mu\,(2 m_\mu)$, the decay branch of $\phi \to e \mu\,(\mu \mu)$ opens, and the decay length of the light boson shortens. The smaller coupling is, therefore, preferred to reach the detector of the E137 experiment.
Since these regions cannot be excluded only by analyzing the $ee$ decay search, the CLFV couplings 
can be further constrained in these regions. Thus, it is important to search for not only the CLFC decays 
but also the CLFV decays as signals in future beam dump experiments.
The FASER experiment \cite{Feng:2017uoz,Feng:2017vli,FASER:2018eoc}, which will start next year, will be able to perform the searches 
for the dark photon decay into $e\mu$ pairs. Theoretical analyses will be shown in \cite{faser-lfv}.

\section{Summary}
\label{sec:summary}
We have studied the constraints from the electron beam dump experiment taking into account 
the charged lepton flavor violating interactions. The scalar, vector, and dipole type interactions 
are considered and the excluded regions have been derived for the search for $X \to ee$. 
We have found that the parameter regions unconstrained by $\mu \to eee$ and $\mu \to e X$ can be excluded 
for three interactions.
The exclusion regions depend on the CLFV and CLFC couplings as well as the boson mass, 
and new bounds can be derived when the ratios of the CLFV coupling to the CLFC one are $\lsim 100$.
For the illustrative purpose, we have also derived the presumed excluded region for the search of 
$\phi \to ee,~ e\mu,~\mu\mu$ decays in the same setup with the E137 experiment. It has been found 
that the excluded regions can be extended and the CLFV couplings can be further constrained in those  
regions. Such searches and analyses for the CLFV decays will be important to new physics searches 
in future experiments.

\begin{acknowledgements}
This work was supported by JSPS KAKENHI Grant Number JP20K04004 [YT], 
~JP18H01210 [TA, TS],~JP18K03651, and MEXT KAKENHI Grant Number JP18H05543 [TS].
\end{acknowledgements}

\appendix
\section{Effective photon flux}
\label{apdx:eff-photon-flux}
The effective photon flux $\xi$ produced by an atom with mass number $A$ and atomic number $Z$ is defined by~\cite{Bjorken:2009mm}
\begin{align}
    \xi = \int_{t_{\mathrm{min}}}^{t_{\mathrm{max}}} dt \frac{t - t_{\mathrm{min}}}{t^2} G_2(t)~,
\end{align}
where $t_{\mathrm{min}} = (m_X^2/2 E_e)^2$ and $t_{\mathrm{max}} = m_X^2$. The electric form factor $G_2(t)$ is given by elastic and 
inelastic components,
\begin{align}
    G_2(t) = G_{2, \mathrm{el}} + G_{2, \mathrm{inel}}~.
\end{align}
The elastic electric form factor is given by 
\begin{align}
    G_{2, \mathrm{el}} = \left( \frac{a^2 t}{1+a^2 t} \right)^2 \left( \frac{1}{1 + \frac{t}{d}} \right)^2 Z^2~,
\end{align}
where $a=111 Z^{-1/3}/m_e$ and $d=0.164~\mathrm{GeV}^2 A^{-2/3}$.
The inelastic electric form factor is given by 
\begin{align}
    G_{2, \mathrm{in}} = \left( \frac{{a'}^2 t}{1+{a'}^2 t} \right)^2 \left( \frac{1 + \frac{t}{4m_p^2} (\mu_p^2 - 1)}{(1 + \frac{t}{0.71~\mathrm{GeV}^2})^4} \right)^2 Z~,
\end{align}
where $a'=773Z^{-2/3}/m_e$, $m_p$ is the proton mass and $\mu_p = 2.79$. 

The electric form factor is dominated by the elastic one in our 
study, thus it scales by $Z^2$. See \cite{Bjorken:2009mm} for 
more detail.

\section{Function $\tilde{U}_n^\ell$}
\label{apdx:tilde-u}
The functions $\tilde{U}_n^\ell~(n=1\mathchar`-4)$ are defined by
\begin{align}
    \tilde{U}_n^\ell = \int^{\theta_\mathrm{max}}_0 \frac{\sin\theta}{[\theta^2 + \eta_\ell(x)]^n} d\theta~,
    \label{eq:tilde-U}
\end{align}
where $\theta_\mathrm{max}$ is the maximal angle determined by angular acceptance of the detector, and $\eta_\ell(x)$ is given in \eqref{eq:eta-l}. For the small angle $\theta_\mathrm{max}$, $\tilde{U}_n^\ell$ can be approximated by the following formulae~:
\begin{subequations}
\begin{align}
    \tilde{U}_1^\ell &\simeq
    \frac{1}{2} \log \left(\frac{\eta_\ell +\theta_{\mathrm{max}}^2}{\eta_\ell}\right)
   +
   \frac{1}{12} \left(\eta_\ell  \log \left(\frac{\eta_\ell +\theta_{\mathrm{max}}^2}{\eta_\ell }\right)-\theta_{\mathrm{max}}^2\right) 
   \nonumber \\
   &\qquad +
    \frac{1}{480} \left(2 \eta_\ell ^2 \log \left(\frac{\eta_\ell +\theta_{\mathrm{max}}^2}{\eta_\ell }\right)-2 \eta_\ell 
   \theta_{\mathrm{max}}^2+\theta_{\mathrm{max}}^4\right)
    , \\
    \tilde{U}_2^\ell &\simeq
    \frac{\theta_{\mathrm{max}}^2}{2 \eta_\ell \left(\eta_\ell +\theta_{\mathrm{max}}^2\right)}
    +
    \frac{1}{12} \left(-\frac{\eta_\ell }{\eta_\ell +\theta_{\mathrm{max}}^2} 
    -\log \left(\frac{\eta_\ell +\theta_{\mathrm{max}}^2}{\eta_\ell}\right) +1\right) 
   \nonumber \\
   &\qquad +
    \frac{1}{120} \left(\frac{(2 \eta_\ell +\theta_{\mathrm{max}}^2) \theta_{\mathrm{max}}^2}{2 \left(\eta_\ell +\theta_{\mathrm{max}}^2\right)}
     - \eta_\ell  \log \left(\frac{\eta_\ell +\theta_{\mathrm{max}}^2}{\eta_\ell }\right)\right)
   ,\\
    \tilde{U}_3^\ell &\simeq
    \frac{(2\eta_\ell + \theta_{\mathrm{max}}^2) \theta_{\mathrm{max}}^2}{4 \eta_\ell^2 \left(\eta_\ell
   +\theta_{\mathrm{max}}^2\right)^2}
   -
   \frac{\theta_{\mathrm{max}}^4}{24 \eta_\ell  \left(\eta_\ell
   +\theta_{\mathrm{max}}^2\right)^2}
   \nonumber \\
   & \qquad +
   \frac{1}{120} \left(\frac{1}{2} \log \left(\frac{\eta_\ell +\theta_{\mathrm{max}}^2}{\eta_\ell
   }\right)-\frac{(2 \eta_\ell +3 \theta_{\mathrm{max}}^2) \theta_{\mathrm{max}}^2}{4 \left(\eta_\ell +\theta_{\mathrm{max}}^2\right)^2}\right)
    ,\\
    \tilde{U}_4^\ell &\simeq
    \frac{(3 \eta_\ell^2 + 3 \eta_\ell \theta_{\mathrm{max}}^2 + \theta_{\mathrm{max}}^4) \theta_{\mathrm{max}}^2}{6 \eta_\ell^3 \left(\eta_\ell +\theta_{\mathrm{max}}^2\right)^3}
    -
    \frac{(3 \eta_\ell +\theta_{\mathrm{max}}^2) \theta_{\mathrm{max}}^4}{72 \eta_\ell ^2 \left(\eta_\ell +\theta_{\mathrm{max}}^2\right)^3}
    +
    \frac{\theta_{\mathrm{max}}^6}{720
   \eta_\ell  \left(\eta_\ell +\theta_{\mathrm{max}}^2\right)^3}
    ~.
\end{align}
\end{subequations}

\section{Widths of CLFV decays of muons}
\label{apdx:clfv-deca-width}
\begin{figure}[t]
\centering
\includegraphics[width=1.0\textwidth]{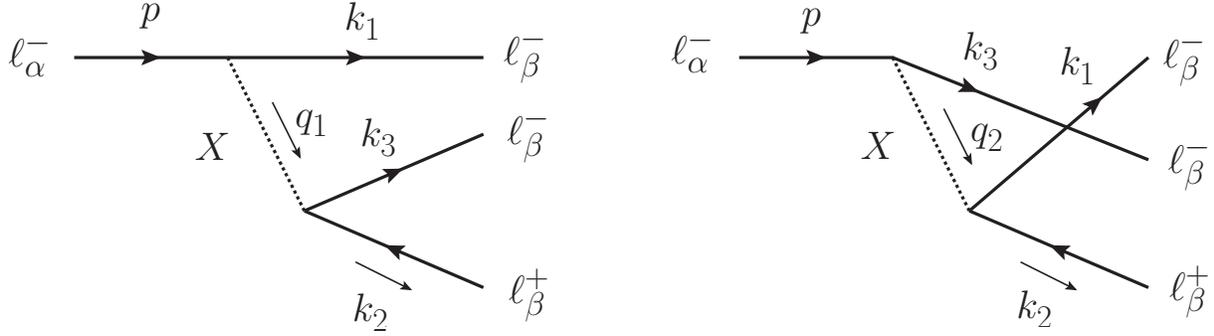}
\caption{
Feynman diagrams of the process $\ell_\alpha^- \rightarrow \ell_\beta^- \ell_\beta^- \ell_\beta^+$, in which $\ell_\alpha^-$ represents the SM charged lepton and $X=\phi, ~Z', ~A'$.
}
\label{fig:m3e}
\end{figure}
We here discuss the experimental limits on the CLFV decays of the muon.
The CLFV is tightly constrained by the rare muon decays, i.e., $\mu \rightarrow e\gamma$ and $\mu \rightarrow eee$.
The current strongest limits on the branching ratio of $\mu \to e \gamma$ is given by MEG~\cite{MEG:2016leq} as BR$(\mu \to e \gamma) < 4.2 \times 10^{-13}$.
On the other hand, the strongest limits of $\mu \to eee$ is given by SINDRUM~\cite{SINDRUM:1987nra} as BR$(\mu \to eee) < 10^{-12}$.
Though, the constraint on BR$(\mu \to e \gamma)$ is stronger than that on BR$(\mu \to eee)$, the decay width of $\mu \rightarrow e\gamma$ is suppressed by the electromagnetic coupling and a loop factor in comparison with that of $\mu \rightarrow eee$.
As a result, usually, the bound on CLFV couplings from $\mu \rightarrow eee$ is stronger than that from $\mu \rightarrow e\gamma$. 
We, therefore, take into account only the constraint from $\mu \rightarrow eee$ for $m_X > m_\mu - m_e$ in this work.
For $m_X < m_\mu - m_e$, muons can decay into the light bosons.
The TWIST collaboration~\cite{TWIST:2014ymv} gives the limits on the branching ratio of $\mu \to e X$ as BR$(\mu \to e X) < 5.8 \times 10^{-5}$ for $m_X = 13\,\mathchar`-\,80$ MeV.
The PIENU collaboration~\cite{PIENU:2020loi} also gives the upper limit as BR$(\mu \to e X) \lesssim 10^{-5}\,\mathchar`-10^{-4}$ for $m_X = 47.8\,\mathchar`-\,95.1$ MeV.
Moreover, Ref.~\cite{Derenzo:1969za} gives the upper limit as BR$(\mu \to e X) < 2 \times 10^{-4}$ for $m_X = 98.1\,\mathchar`-\,103.5$ MeV.
These limits give much more severe constraints on the CLFV coupling than that from the electron beam dump, and then we focus on the region where the light boson is heavier than $m_\mu - m_e$.

The partial decay width of $\ell_\alpha^- \rightarrow \ell_\beta^- \ell_\beta^- \ell_\beta^+$ is given by
\begin{align}
\Gamma = \frac{1}{(8 \pi m_\alpha)^3}
\int^{s_{12}^{\mathrm{max}}}_{s_{12}^{\mathrm{min}}} \!\! ds_{12} 
\int^{s_{23}^{\mathrm{max}}}_{s_{23}^{\mathrm{min}}}  \!\! ds_{23}~ 
\frac{1}{2} \sum_{\rm spin} |M|^2~,
\end{align}
where $s_{12}=(p-k_3)^2$ and $s_{23}=(p-k_1)^2$. 
The integral ranges of $s_{12}$ and $s_{23}$ are given by
\begin{subequations}
\begin{align}
s_{23}^{\mathrm{max}} &=  (\hat{E}_2 + \hat{E}_3)^2 - (\hat{P}_2 - \hat{P}_3)^2~,\\
s_{23}^{\mathrm{min}} &=   (\hat{E}_2 + \hat{E}_3)^2 - (\hat{P}_2 + \hat{P}_3)^2~,\\
s_{12}^{\mathrm{max}} &=  (m_\alpha - m_\beta)^2~,\\
s_{12}^{\mathrm{min}} &= 4 m_\beta^2~,
\end{align}
\end{subequations}
where
\begin{subequations}
\begin{align}
\hat{E}_2 &= \frac{1}{2 \sqrt{s_{12}}}  ( m_\alpha^2 - s_{12} - m_\beta^2 )~, \\
\hat{E}_3 &=  \frac{\sqrt{s_{12}}}{2}~, \\
\hat{P}_2 &= \frac{1}{2} \sqrt{ s_{12} - 2 (m_\alpha^2 + m_\beta^2) + \frac{(m_\alpha^2 - m_\beta^2)^2}{s_{12}} }~, \\
\hat{P}_3 &=  \frac{1}{2} \sqrt{ s_{12} - 4 m_\beta^2}~.
\end{align}
\end{subequations}
The amplitude is written as
\begin{align}
\frac{1}{2}\sum_{\rm spin} |{\cal M}|^2 = 
\frac{C^2}{2}\left[ ~P_X^2(s_{23})~M^2_1 + P_X^2(s_{12})~M^2_2 - 2P_X(s_{23})P_X(s_{12})~{\rm Re}[M^2_3]~ \right]~,
\end{align}
where $P_X\left(s_{23(12)}\right)= \left(s_{23(12)}-m_X^2\right)^{-1}$, 
$X = \phi, Z', A'$, and $C = y_{e}y_{e\mu}', ~g's^3c$ and $\mu_e \mu'$ for the scalar, the vector and the dipole interaction, respectively.
For the scalar interaction,
\begin{subequations}
\begin{align}
M^2_1 &= -4(s_{23}-4m_\beta^2)(s_{23}-(m_\alpha + m_\beta)^2)~,\\
M^2_2 &= -4(s_{12}-4m_\beta^2)(s_{12}-(m_\alpha + m_\beta)^2)~,\\
M^2_3 &= 2s_{12}s_{23} + 4m_\beta(m_\alpha + m_\beta)(s_{12}+s_{23})\nonumber\\
&\hspace{10mm}-2m_\beta(m_\alpha + m_\beta)(5m_\beta^5 + 2m_\alpha m_\beta + m_\alpha^2)~,
\end{align}
\end{subequations}
for the vector interaction,
\begin{subequations}
\begin{align}
    M_1^2 &= -8 \big\{ 2s_{12}^2 + s_{23}^2 + 2s_{12} s_{23}
     - (m_\alpha - m_\beta)^2 s_{23} - 2 (m_\alpha^2 + 3 m_\beta^2) s_{12} \nonumber \\
    &\qquad + 2 m_\beta^2 (m_\alpha + m_\beta)^2
     \big\}~, \\
    M_2^2 &= -8 \big\{ 2s_{23}^2 + s_{12}^2 + 2s_{12} s_{23}
     - (m_\alpha - m_\beta)^2 s_{12} - 2 (m_\alpha^2 + 3 m_\beta^2) s_{23} \nonumber \\
    &\qquad + 2 m_\beta^2 (m_\alpha + m_\beta)^2
     \big\}~, \\
    M_3^2 &= 8 (s_{12} + s_{23}+m_\beta(m_\alpha + m_\beta)) 
        (s_{12} + s_{23} - m_\alpha^2 + m_\alpha m_\beta -2 m_\beta^2)~,
\end{align}
\end{subequations}
and for the dipole interaction,
\begin{subequations}
\begin{align}
M^2_1 &= -4s_{23}
\left\{
s_{23}^3 + 4(s_{12} + s_{23})s_{12}s_{23} 
- (m_\alpha - m_\beta)^2s_{23}^2 
\right. \nonumber\\
&\hspace{10mm}\left.
- 4(m_\alpha^2 + 3m_\beta^2)s_{12}s_{23} 
+ 8m_\beta^2(m_\alpha + m_\beta)^2 s_{23}
- 4m_\beta^2 (m_\alpha^2 - m_\beta^2)^2
\right\}~, \\
M^2_2 &= -4s_{12}
\left\{
s_{12}^3 + 4(s_{12} + s_{23})s_{12}s_{23} 
- (m_\alpha - m_\beta)^2s_{12}^2 
\right. \nonumber\\
&\hspace{10mm}\left.
- 4(m_\alpha^2 + 3m_\beta^2)s_{12}s_{23} 
+ 8m_\beta^2(m_\alpha + m_\beta)^2 s_{12}
- 4m_\beta^2 (m_\alpha^2 - m_\beta^2)^2
\right\}~, \\
M^2_3 &= 2
\left\{
2(s_{12}^2 + s_{23}^2)s_{12}s_{23} + 5s_{12}^2s_{23}^2
- 2(m_\alpha^2 + 3m_\beta^2)(s_{12}+s_{23})s_{12}s_{23}
\right. \nonumber\\
&\hspace{10mm}\left.
+ m_\beta(m_\alpha+m_\beta)(3m_\alpha^2 + 2m_\alpha m_\beta + 11m_\beta^2)s_{12}s_{23}
\right. \nonumber\\
&\hspace{10mm}\left.
- 2m_\beta^2(m_\alpha^2 - m_\beta^2)^2(s_{12}+s_{23})
- 4m_\beta^3(m_\alpha + m_\beta)(m_\alpha^2 - m_\beta^2)^2
\right\}~,
\end{align}
\end{subequations}
where $m_\alpha$ stands for the mass of the $\alpha$ flavor charged lepton.

The partial width of the muon decay into an electron and a light boson is given by
\begin{align}
\Gamma(\mu \rightarrow e \phi)
= &\,\frac{\lambda_{e\mu}^2}{16\pi}m_\mu~
\lambda\left( \frac{m_e^2}{m_\mu^2},\frac{m_\phi^2}{m_\mu^2} \right)
\left\{ 1+2\frac{m_e}{m_\mu}+\frac{m_e^2 - m_\phi^2}{m_\mu^2} \right\}~, \\
\Gamma(\mu \rightarrow e Z')
= &\,\frac{(g'sc)^2}{16\pi}m_\mu~
\lambda\left( \frac{m_e^2}{m_\mu^2},\frac{m_{Z'}^2}{m_\mu^2} \right)
\frac{m_\mu^2}{m_{Z'}^2} \nonumber \\
&\times \left\{
\left( 1-\frac{m_e^2}{m_\mu^2} \right)^2 - \frac{m_{Z'}^4}{m_\mu^4}
+\left( 
1-6\frac{m_e}{m_\mu}+\frac{m_e^2 - m_{Z'}^2}{m_\mu^2} 
\right)\frac{m_{Z'}^2}{m_\mu^2}
\right\}~, \\ 
\Gamma(\mu \rightarrow e A')
= &\,\frac{(\mu' m_\mu)^2}{16\pi}m_\mu~
\lambda\left( \frac{m_e^2}{m_\mu^2},\frac{m_{A'}^2}{m_\mu^2} \right) \nonumber \\
&\times \left[
2\left\{
\left( 1-\frac{m_e^2}{m_\mu^2} \right)^2 - \frac{m_{A'}^4}{m_\mu^4}
\right\}
-\left( 
1+6\frac{m_e}{m_\mu}+\frac{m_e^2 - m_{A'}^2}{m_\mu^2} 
\right)\frac{m_{A'}^2}{m_\mu^2}
\right]~,
\end{align}
for the scalar, vector, and dipole interaction, respectively, and $\lambda(a,b)$ is the Kallen function given in Eq. (\ref{eq:kallen}).

\bibliographystyle{apsrev}
\bibliography{biblio}

\end{document}